\documentstyle[preprint,prl,aps]{revtex}
\begin{document}
\title{ 
SU(2,1) Dynamics of Multiple Giant Dipole Resonance Coulomb 
Excitation}
\author{ 
{\it M.S. Hussein, A.F.R. de Toledo Piza, O.K. Vorov
}}
\address{ 
Instituto de Fisica, 
Universidade de Sao Paulo \\
Caixa Postal 66318,  05315-970,  \\
Sao Paulo, SP, Brasil 
}
\date{13 August 1999}
\maketitle
\begin{abstract}
We construct a three-dimensional analytically
soluble model of the nonlinear effects in 
Coulomb excitation of 
multiphonon Giant Dipole Resonances (GDR)
based on the SU(2,1) algebra. 
The full 3-dimensional model predicts further 
enhancement of the Double GDR (DGDR) cross sections at high 
bombarding energies.
Enhancement factors for DGDR measured in 
thirteen different processes
with various projectiles and targets
at different bombarding energies are well reproduced with 
the same value
of the nonlinearity parameter
with the exception of the anomalous case of $^{136}$Xe
which requires a larger value.
%
%
%
%
%
%
%
\end{abstract}
\newpage
One of the most interesting applications of Coulomb Excitation
in heavy ion collisions 
\cite{book,EMLING,ABE,BCHTP,BP,bbabs,CH-FR,SCH-EXP-Xe}
is investigation of
multi phonon nuclear Giant Resonances (GR) \cite{EMLING}.
Possibility to excite multiple GR 
involves Bose statistics
of collective excitations and constitutes 
the ``family property'' 
of vibrational collective motion in both finite 
(nuclei, clusters) and infinite quantum systems.
Within this concept, making no distinction between
infinite and finite systems, the excitation process 
can be modeled via
a multidimensional quantum oscillator coupled linearly 
to the external
time-dependent field, providing excellent agreement with the 
single-GR experimental data \cite{EMLING}.

Validity of this completely linear theory has been questioned  
by nearly all the experimental data wherever multi-phonon GR
has been observed\cite{EMLING}: 
the double Giant Dipole Resonance (GDR) excitation cross sections
\cite{SCH-EXP-Xe}
are found about $1.3-2$ bigger than follows from theory.
This shortcoming of the linear theory, known as 
``enhancement factor problem'', has been addressed widely 
in current literature
within a number of approaches:
higher order perturbation theory\cite{BZ93}, anharmonic effects
\cite{CH-FR},\cite{B-D}, 
concept of hot phonons
\cite{hot-phonon1,hot-phonon2,nov16}. 
Clearly, nonlinear effects, that in principle 
can not be neglected in 
finite
Fermi
systems \cite{BZ}, are not easily dealt with 
either at a microscopical
or  even at a phenomenological level  \cite{CH-FR}.

It is therefore appealing to construct 
a natural, soluble ``minimal extension''
of the 
harmonic model of Coulomb excitations, in which deviations from
the linear scheme are reasonably described via a few parameters. 
Without dealing in depth with the microscopic theory,
we present here such a ``minimal extension''.
This single-parameter nonlinear model can be solved
exactly
using algebraic properties of boson operator combinations
forming algebra 
SU(2,1).
The model allows us to correlate all the experimental data
for the ``enhancement factors'' in various nuclei and various 
bombarding energies, using a single value of the 
universal nonlinear parameter.

Within the
semiclassical approach \cite{AW65} to 
Coulomb excitation,
the projectile motion 
is approximated by a constant velocity $v$ on a 
straight line classical trajectory 
with impact parameter $b$
and
internal excitation 
is treated 
quantum mechanically. 
The intrinsic state $|$$\Psi$$($$t$$)$$\rangle$ of the system
undergoing excitation 
obeys
the time-dependent Schr\"odinger equation,
\begin{equation} \label{SCHR}
i {\partial /\partial t} \vert\Psi(t)\rangle   =
\left[ H_0 + V (t)  \right]\ \vert\Psi(t)\rangle,
\qquad  \vert\Psi(t=-\infty)\rangle = \vert 0 \rangle,
\end{equation}
where  $H_0$ is the intrinsic Hamiltonian and 
$V(t)=v_1(t)[{\cal D}_{-1}^{\dagger}-{\cal D}_{+1}^{\dagger}]+
v_0(t){\cal D}_{0}^{\dagger}+h.c.$
is the channel-coupling interaction
with
${\cal D}^{\dagger}$ and ${\cal D}$
the 
dipole operators acting in the
space of the multi-GDR states created by the boson operators
$d^{+}_m$,
with the angular momentum projection $m$ ($\hbar$$=$$c$$=1$).
The functions $v_{1(0)}(t)$ 
describe the interaction with the electromagnetic field
\cite{EMLING},\cite{BCHTP} 
(see below).
The excitation probability of an intrinsic
state $\vert N\rangle$ with N phonons in a collision with 
impact parameter $b$  
and the 
total cross section $\sigma_N$ are 
\cite{EMLING,BCHTP,BP,AW65}:
\begin{equation} \label{Pn}
P_N(b) 
= | \langle N | \Psi(t=\infty) \rangle |^2, \qquad
\sigma_N= 2 \pi 
\int^{\infty} b P_N(b).
\end{equation}

The internal nuclear Hamiltonian $H_0$ is nearly harmonic 
\cite{EMLING},\cite{CCG},\cite{Bertsch-Feldmeier}, so 
$H_0 = \omega N=\omega \sum_m d^{\dagger}_m d_m$.
In principle, 
this does not exclude anharmonicities \cite{PIZA} in 
the transition operators 
${\cal D}^{\dagger}$, ${\cal D}$
when expanded in terms of phonon operators
\begin{equation} \label{E1} 
{\cal D}^{\dagger}_m
=d^{\dagger}_m +  
x  \sum_{m_1} d^{\dagger}_{m} 
d^{\dagger}_{m_1} d_{m_1} + 
\sum_{m_1} x_{m_1} d^{\dagger}_{m} d^{\dagger}_{m_1} 
d^{\dagger}_{m_1} + 
x_2 \sum_{m_1 m_2} d^{\dagger}_{m}
d^{\dagger}_{m_1} d_{m_1}d^{\dagger}_{m_2} d_{m_2} + ...,
\end{equation}
These effective nonlinearities 
\cite{BZ},\cite{ABE}
could result from
perturbation theory treatment of anharmonicities in the phonon
Hamiltonian,
from coupling to other degrees of freedom, both collective
(e.g., quadrupole GR) and noncollective \cite{hot-phonon1} etc.
The linear limit of the problem, 
${\cal D}^{\dagger}_m=d^{\dagger}_m$,
is exactly soluble
giving the Poisson formula for the excitation probabilities 
\begin{eqnarray} \label{POISSON}
P_N = e^{-\rho} \frac{\rho^{N} }{N!},
\qquad \qquad
\rho=\sum_{m=0,\pm 1} |\alpha^{harm}_m|^2=
\sum_{m=0,\pm 1} 
\vert \int_{-\infty}^{\infty}v_m(t)e^{i\omega t} dt \vert^2
\end{eqnarray}
where the amplitudes $\alpha^{harm}_m$ 
are given by the 
modified Bessel functions $K_1$ and $K_0$ \cite{EMLING}. 

In order to reduce the number of unknown parameters
in Eq.(\ref{E1})
it is reasonable to restrict the 
higher-order nonlinear 
corrections 
$\propto$$x_i$. 
We save the first, dominating, 
nonlinear 
term in (\ref{E1}) with its coefficient $0$$\leq$$x$$\ll$$1$ 
and save only those terms in (\ref{E1}) which match 
terms appearing in the expansion of the square root,
${\cal D}^{\dagger}_m\rightarrow d^{\dagger}_m 
(1+2x \sum_m d^+_m d_m)^{1/2}$.

The nonlinear effects are now controlled 
by the single parameter $x$.
This may be justified, by noting that 
the leading term is saved, while the ansatz for the higher order
terms, which should be small anyway, 
obeys the basic requirement that they get smaller 
as the order increases.
Further, this parametrization leads to a soluble
problem.

The group theoretical solution is based 
on the consideration of
the eight operators 
\begin{eqnarray} \label{SU21-SHORT}
D^+ = (d^+_{-1} - d^+_{+1}  )\sqrt{k + N/2}, \quad
D^+_0 = d^+_{0}\sqrt{ 2k + N}, \quad
J^+ = \frac{1}{2^{1/2}} \left( d^+_{-1} -d^+_{+1} \right) d_0 ,
\nonumber\\
D^0 = \frac{1}{4} 
\left[ (d^+_{-1} -  d^+_{+1} )(d_{-1} - d_{+1} ) + 
2(2k+N) \right], \quad
D^0_0 = \frac{1}{2} 
\left[2k+N +d^+_0 d_0\right].
\end{eqnarray}
and 
$D$$^-$$=$$($$D$$^+$$)$$^{\dagger}$, 
$D$$^-_0$$=$$($$D$$^+_0$$)$$^{\dagger}$,$J$$^-$$=$
$($$J$$^+$$)$$^{\dagger}$
with $k$$\equiv$$\frac{1}{4x}$ the Casimir invariant.
They form a closed SU(2,1) algebra, the non-compact 
analogue of SU(3).
The pairwise commutators between
(\ref{SU21-SHORT}) can be evaluated directly: 
Three of them are 
$[$$D$$^-$$,$$D$$^+$$]$$=$$2$$D$$^0$,
$[$$D$$^-_0$$,$$D$$^+_0$$]$$=$$2$$D$$^0_0$,
$[$$J$$^-$$,$$J$$^+$$]$$=$$2$$($$D$$^0_0$$-$$D$$^0$$)$.
The other nonzero ones are
$[$$D$$^-$$,$$D$$^0$$]$$=$$2$$[$$D$$^-$$,$$D$$^0_0$$]$$=$
$[$$D$$^-_0$$,$$J$$^-$$]$$=$$D$$^-$; 
$\quad$$
$$[$$D$$^-_0$$,$$D$$^0_0$$]$$=$$2$$[$$D$$^-_0$$,$$D$$^0$$]$$=$
$[$$D$$^-$$,$$J$$^+$$]$$=$$D$$^-_0$, 
and 
$[$$D$$^-$$,$$D$$^+_0$$]$$=$$2$$[$$J$$^-$$,$$D$$^0$$]$$=$
$-$$2$$[$$J$$^-$$,$$D$$^0_0$$]$$=$$J$$^-$. 
The remaining nonzero commutators
are given by Hermitean conjugates to the above.
In the interaction representation, the evolution equation 
$i\frac{\partial }{\partial t}
|\psi(t)\rangle=e^{iH_0 t}V(t)e^{-iH_0 t}|\psi(t)\rangle$
(\ref{SCHR}) with (\ref{E1}) 
reads:
\begin{eqnarray} \label{INTERACTION}
i(\partial/\partial t)|\psi(t)\rangle=
2 x^{1/2}
\left[ \tilde{v}_1(t) D^{\dagger} + 
\tilde{v}^*_1(t) D^-  
+
\tilde{v}_0(t)   D^{\dagger}_0 + 
\tilde{v}^*_0(t) D^-_0  \right] |\psi(t)\rangle.
\end{eqnarray}
where ${\tilde v}_1\equiv v_1 e^{i\omega t}$
and ${\tilde v}_0\equiv v_0 e^{i\omega t}/\sqrt{2}$.
Any product of exponentials involving the operators from the set
(\ref{SU21-SHORT}) can be reduced to a simpler exponential
(see, e.g.,\cite{GWI}),
as due to closure of the pairwise commutators 
between (\ref{SU21-SHORT}),
no new operator structures arise 
while re-arranging order of the operators in the time-ordered
exponential (see, e.g.,\cite{KLEINERT})
that solves (\ref{SCHR}),(\ref{INTERACTION})
\begin{equation}  \label{DECOMPOSITION}
|\psi(t)\rangle \equiv
T exp\left\{ 
\int_{-\infty}^{t}e^{iH_0 \tau}V(\tau)e^{-iH_0 \tau}d\tau
\right\}|0\rangle
=
e^{ a D^+ + b D^+_0 } e^{ f J^+}  e^{ c D^0 + d D^0_0 } e^{ g J^-}
e^{ a' D^- + b' D^-_0 } |0\rangle 
\end{equation}
where the eight time-dependent c-numbers, the Latins $a-g$, must be
chosen so $\psi(t)\rangle$ (\ref{DECOMPOSITION}) 
obeys the  Schr\"odinger 
equation 
(\ref{INTERACTION}).
The 
ansatz
(\ref{DECOMPOSITION}) can be simplified using 
$D^-|0\rangle=D^-_0|0\rangle=J^{\pm}|0\rangle=0;
D^0|0\rangle=D^0_0|0\rangle=k|0\rangle$,
which follow from (\ref{SU21-SHORT}).
The expression (\ref{DECOMPOSITION}) then reduces to
\begin{equation}\label{DECOMPOSITION-2}
|\psi(t)\rangle = 
\left[ 1 - 4x\left( |\alpha(t)|^2 + \beta(t)|^2\right)
\right]^{\frac{1}{4x}}
e^{i \phi(t)}
e^{ \frac{\alpha(t)}{\sqrt{k}} D^+ + 
\frac{\beta(t)}{\sqrt{k}} D^+_0 } |0\rangle,
\end{equation} 
where the first factor comes from unitarity,
$\langle\psi(t)|\psi(t)\rangle=1$.
The phase $\phi(t)$ is unimportant for the following.
Substituting (\ref{DECOMPOSITION-2})
into 
(\ref{INTERACTION}), we obtain, after some lengthy
algebra using heavily the operator identity
$
e^{A}Be^{-A}=B + [A,B] + (1/2!)[A,[A,B]] + 
...$,
the two nonlinear equations for 
the amplitudes $\alpha$ and $\beta$
\begin{eqnarray}   \label{RICCATI}    
i\partial\alpha/\partial t-Q\alpha={\tilde v}_1(t),
\quad  
i\partial\beta/\partial t-Q\beta  = {\tilde v}_0(t),
\end{eqnarray}
with 
$Q\equiv4x[{\tilde v}_0^*(t) \beta + {\tilde v}_1^*(t)\alpha]$
and the initial condition $\alpha(-\infty)=\beta(-\infty)=0$.
After projecting the asymptotic 
state (\ref{DECOMPOSITION-2}) at $t=\infty$ onto the states
with definite number of GDR phonons, we obtain
the 
non-Poissonian
expression for the excitation probabilities
\begin{eqnarray} \label{RESULT}
\quad  P_{N} = 
\frac{ \Gamma(\frac{1}{2x}+N) \left[4x\left(|\bar{\alpha}|^2 + 
|\bar{\beta}|^2\right)
\right]^{N}}
{ N! \Gamma(\frac{1}{2x})
\left[ 1 - 4x\left( |\bar{\alpha}|^2 +  \bar{\beta}|^2\right)
\right]^{-\frac{1}{2x}} 
}.
\end{eqnarray}
where $\bar{\alpha}$ and $\bar{\beta}$ are 
the asymptotic solutions
of the system (\ref{RICCATI}) at $t=\infty$. Their values can be 
easily tabulated by solving (\ref{RICCATI}).  
If $x\rightarrow 0$, Eq.(\ref{RESULT}) is reduced to Poisson,
while the solutions to (\ref{RICCATI}) are reduced to the
harmonic amplitudes (\ref{POISSON}),
$|\alpha|\rightarrow|\alpha^{harm}_1|=
\frac{2F\xi^2}{\omega}K_1(\xi)$ and   
$|\beta|\rightarrow|\alpha^{harm}_0|/\sqrt{2}=
\frac{2F\xi^2}{\gamma\omega}K_0(\xi)$. 
Thus, the
harmonic results are restored.
At nonzero nonlinearity $x>0$, the multiple GDR excitation 
probabilities $P_{N}$ 
turn out to be enhanced as compared to their 
values $P^{harm}_{N}$ in the harmonic limit, Eq.(\ref{POISSON}).
The cross sections $\sigma_N$ and their harmonic 
values,$\sigma_N^{harm}$ 
are given by integrating in Eq.(\ref{Pn}) from 
the grazing value $1.2(A^{1/3}_{ex}+A^{1/3}_{sp})$.
The enhancement factors $r_N=\sigma_N/\sigma_N^{harm}$
can be studied and compared with the
data.
The functions $v_m(t)$, in the 
notations
$\xi=\frac{\omega b}{v\gamma}$, $\tau=\frac{\omega t}{\xi}$
are \cite{BCHTP,BP}
\begin{eqnarray} \label{v1short}
v_{1}(t) = Ff^3, 
\quad
v_0(t) = F\sqrt{2}\gamma\left[-\frac{\partial f}{\partial\tau}-
i \xi v^2 f\right], \quad 
F=\frac{Z_{sp} e^2 \gamma}{2 b^2}
\left[
\frac{ N_{ex} Z_{ex} }{A_{ex}^{2/3} m_N \cdot 80 MeV} 
\right]^{\frac{1}{2}}.
\end{eqnarray}
where $f=(1+\tau^2)^{-\frac{1}{2}}$
and $\gamma=(1-v^2)^{-\frac{1}{2}}$. 
In the strength $F$,
$m_N$ and $e$ are the proton mass and charge,
$Z$, $N$ and $A$ denote the nuclear charge,
the neutron number and the mass numbers,
the labels $ex$ ($sp$) refer 
to the excited (spectator) 
nucleus in the colliding pair.

Let us discuss the energy dependence of the two-phonon enhancement
factor $r_2$. 
In Ref.\cite{HPV1}, we have studied this behavior
for the Pb+Pb system
at the bombarding energies in range $70-700$MeV$\cdot$A, using
truncated (two-dimensional) dynamics and neglecting the 
longitudinal response in the excitation process ($v_{0} \equiv 0$).
Within this ``toy model'' \cite{HPV1}, the dynamics is 
described by the
three operators $D$,$D^+$ and $D^{0}$ which form 
the SU(1,1) subalgebra of the SU(2,1).
The enhancement factor
drops 
steadily
as the bombarding energy grows.
This is not the case {\it at higher energies}, when 
the transverse approximation breaks down
and full solution based on the SU(2,1) algebra is required.
This is illustrated in Fig.1a, where the enhancement factor $r_2$
is shown as a function of 
$\gamma$
for the case of
$Pb + Pb$ collision in the energy range up to 4 Mev,
as compared to the ``transverse approximation''.

The remarkable fact about the full 3-dimensional model
is that the enhancement factor $r_N$ starts to
grow again as the relativistic factor $\gamma$ passes 
the ``extremal'' point 
($\sim 1.3-1.5$ for heavy colliding nuclei). 
Thus the 3-dimensional SU(2,1) model
predicts a new interesting qualitative effect which can 
be tested in experiments using higher bombarding energies.
This 
behavior of $r_N$ is 
related to the
$\gamma$-dependence
of the ``scaled'' longitudinal function ${\tilde v}_{0}$ 
(Eq.\ref{v1short}) \cite{BCHTP,BP}:
${\tilde v}_0(t) =v_0(t)e^{i\omega t}=F\sqrt{2}\left[
-\gamma \frac{d }{d\tau} \left( fe^{i \xi \tau} \right) +
i \frac{\xi}{\gamma}f e^{i \xi \tau}\right]$.
The $\gamma$-dependence of the two terms here is very different.
While the first term scales as $\gamma$,
the absolute maximum of the second one scales as 
$\xi/\gamma =\gamma^{-2}$. 
The first term is proportional to the time derivative 
of the function
$\frac{e^{i\omega t}}{(1+\tau^2)^{3/2}}$
which dies out at 
$t$$=$$\pm$$\infty$.
Therefore, {\it in the  harmonic} limit,
when the longitudinal amplitude 
$\beta$ 
reduces to the simple integral over
time of the function ${\tilde v}_0$, 
the first term simply vanishes. 
The harmonic solution is then given by the integral
of the second term to give $\propto \frac{1}{\gamma}K_0(\xi)$ 
which dies out as $1/\gamma$
as bombarding energy grows, and the longitudinal response
becomes negligible at high bombarding energies
as is well known \cite{EMLING,ABE,BCHTP}. 
This is not the case 
{\it at nonzero nonlinearity} $x \neq 0$: the amplitude $\beta$ 
is given by the solution to the coupled nonlinear
system (\ref{RICCATI}), and the first term in ${\tilde v}_0$ 
not only does contribute but in fact becomes dominant. 
Unlike the second term, the first one grows 
with $\gamma$ and this determines the behavior
of the enhancement factor.

We present below in Figs.1b, Fig.2 and Fig.3a
exact numerical results for the cross sections
calculated according to Eqs.(\ref{Pn}), and (\ref{RESULT}) 
and using numerical solution of Eq.(\ref{RICCATI}).
Since
the nonlinear parameter $x$ is an internal property
of the nucleus in which the GDR is excited, it is reasonable
to expect that it varies from one nuclear species to another.
For the sake of comparison with the experimental data,
it is expedient however to choose a single average value 
$x=0.29$
for all cases and use it to calculate the cross sections.
As seen in Figs.1a,2 and Fig.3a, the nonlinear model can 
in this way (no fitting) 
reproduce
rather well the experimental values 
\cite{SCH-EXP-Xe}
of $r_2$ for 
the twelve excitation
processes
of Pb (Fig.2a,b) and Au (Fig.2c), using different bombarding
energies and targets/projectiles. 
An exceptional case is Xe, where
considerably larger value of $x$ is required.
To illustrate this, the optimal values of $x$ for 
each individual datum are shown in
Fig.3b, together with the adopted average value.

In conclusion,
we presented a simple soluble model to account for the 
nonlinear effects in the transition operators for 
the Coulomb excitation 
of multi-phonon 
GDR
via relativistic heavy ion collisions. 
The solubility of the model is based on the 
group theoretical properties of the boson operators. It allows 
to construct the solution for the dynamics of the multi-phonon 
excitation within the coupled-channel approach.
The well known 
harmonic phonon model 
appears to be a limiting case of the present model when the 
nonlinearity 
goes to zero. 
The main advantages of the limiting harmonic case (unrestricted
multiphonon basis, preservation of unitarity and possibility of
analytical treatment) remain present in our nonlinear
scheme.
Therefore, the model can be viewed as a natural extension of
the harmonic phonon model to include the nonlinear effects in 
a consistent way while keeping the model solvable.
With proper modifications, it can be used in other
(nuclear, molecule {\it etc.}) problems.

At low enough bombarding energies, 
the enhancement factor drops as the bombarding
energy grows. 
This is consistent with the data and 
gives results similar to those recently obtained in a 
different context, with a theory based on the concept 
of fluctuations
(damping) and the Brink-Axel mechanism 
\cite{hot-phonon1},\cite{hot-phonon2}.
The interesting property of the full three-dimensional results 
obtained here is that the 
enhancement factor starts growing again 
at high bombarding energies 
($\sim$ $0.3$$-$$0.5$GeV$\cdot$ A for
heavy colliding nuclei.)
Besides being
an interesting prediction to be tested in
experiment,
this behavior allows one to 
separate the effects of nonlinearity considered here and
the effects proposed in works 
\cite{hot-phonon1,hot-phonon2,nov16}.

The work has been supported by the FAPESP and by the CNPq.

\newpage
\begin{center}
Figure Captions
\end{center}
Fig.1. 
(a) The enhancement factor $r_2$ as a function of
$\gamma$
for $^{208}$Pb$+$$^{208}$Pb collision. 
The 3D results vs.
the transverse approximation results (SU(1,1) toy model, 
\cite{HPV1}) for $x=0.19$.

(b) Enhancement factor $r_2=\sigma_2/\sigma_2^{harm}$
for the DGDR excitation in 
$^{208}Pb$ 
as a function of 
$\gamma$ 
(open symbols connected by a line). 
The data \cite{SCH-EXP-Xe} (filled symbols):
triangle up - Zn projectile,
triangle right - U tagret,
circles - Pb target and projectile. 
{\bf The value of the nonlinear parameter is kept fixed}, 
$x = 0.29$. 

Fig.2.
(a) The same as in Fig.1b, $x=0.29$. Square - Ar projectile, 
diamond -
Kr projectile, triangle down - Ho, and circle - Sn. 
(b) The same for the DGDR in Au. Circle - Kr projectile,
square - Au projectile, diamond - Bi projectile and
triangle - Ne projectile. 
Open circles - theoretical values
for the Bi projectile (the results for Kr and Au projectiles
are the same at $\gamma\sim 2$).

Fig.3.
(a) The same as in Fig.1b,Fig2, but for the DGDR in $Xe$, 
$x=0.29$. 

(b)
The optimal values of the nonlinear parameter $x$ for the
thirteen processes 
vs
strength parameter $F/\omega$, 
Eq.(\ref{v1short}).
The processes are numbered in the order they appear in
Figs.1b, 2a, 2b, 3a (from left to right).

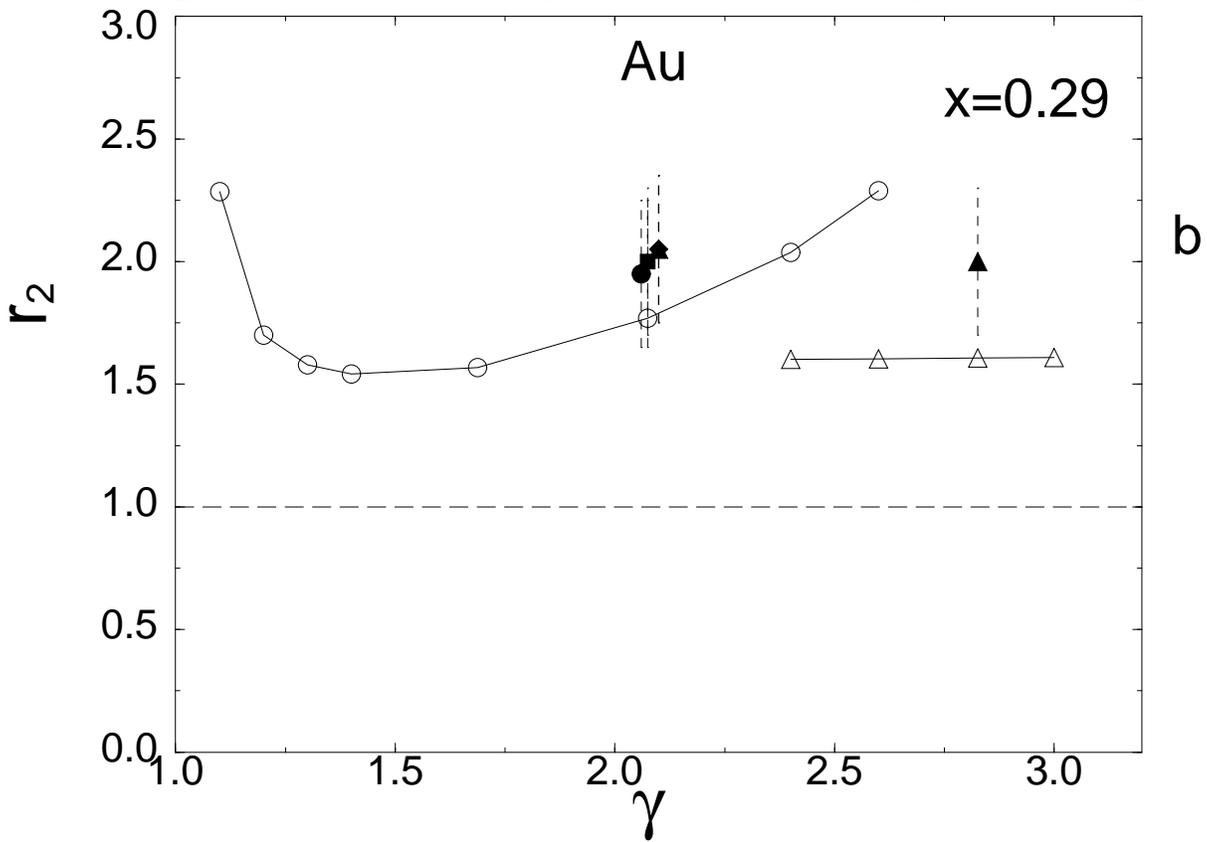
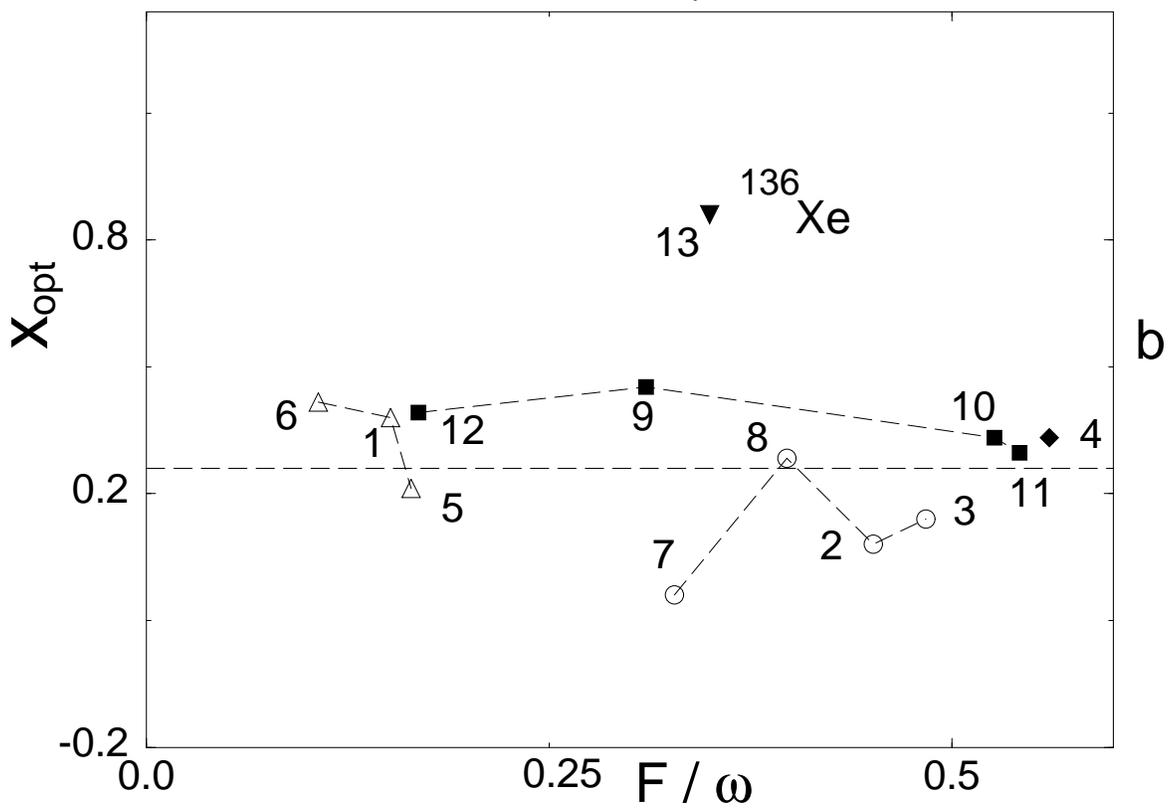
\end{document}